\documentclass[a4paper]{article}
\usepackage{listings}
\usepackage{xcolor}
\usepackage{multirow}
\usepackage{jheppub}  
\usepackage{dcolumn}
\usepackage[english]{babel}
\usepackage[utf8x]{inputenc}
\usepackage[T1]{fontenc}
\usepackage{palatino}
\usepackage{amsmath}
\usepackage{float}
\usepackage{afterpage}
\usepackage{graphicx, subcaption}
\usepackage[colorinlistoftodos]{todonotes}
\usepackage[colorlinks=true]{hyperref}
\usepackage{makeidx}
\newcommand{\be}{\begin{equation}}  
\newcommand{\ee}{\end{equation}}  
\newcommand{\bea}{\begin{eqnarray}}  
\newcommand{\eea}{\end{eqnarray}}  
\makeindex

\makeatletter
\g@addto@macro\bfseries{\boldmath}
\makeatother

\begin{document}

\vspace*{1.2cm}

\thispagestyle{empty}
\begin{center}
{\LARGE \bf Precision measurements of jet production at the ATLAS experiment}

\par\vspace*{7mm}\par

{

\bigskip

\large \bf Francesco Giuli (on behalf of the ATLAS Collaboration\footnote{Copyright 2021 CERN for the benefit of the ATLAS Collaboration. Reproduction of this article or parts of it is allowed as specified in the CC-BY-4.0 license.})}

\bigskip

{\large \bf  E-Mail: francesco.giuli@cern.ch}

\bigskip

{Department of Physics, University of Rome ``Tor Vergata'' and INFN, Section of Rome 2,\\
Via della Ricerca Scientifica 1, 00133, Rome, Italy}

\bigskip

{\it Presented at the Low-$x$ Workshop, Elba Island, Italy, September 27--October 1 2021}

\vspace*{15mm}

\end{center}
\vspace*{1mm}

\begin{abstract}

Measurements of jet production are sensitive to the strong coupling constant, high order perturbative calculations and parton distribution functions. In this talk we present the most recent ATLAS measurements in this area at a centre-of-mass energy of $\sqrt{s}$ = 13~TeV. We present measurements of variables probing the properties of the multijet energy flow and of the Lund Plane using charged particles. We will also present new measurements sensitive to the strong coupling constant. For jet fragmentation, we present a measurement of the fragmentation properties of $b$-quark initiated jets, studied using charged B mesons. This analysis provides key measurements with which to better understand the fragmentation functions of  heavy quarks. All results are corrected for detector effects and compared to several Monte Carlo predictions with different parton shower and hadronisation models.
\end{abstract}
 
 \section{Introduction}
In this proceeding, a review of the most recent ATLAS measurements on jet production is reported. Four different analyses are presented: the first one refers to a measurement of soft-drop jet observables~\cite{ATLAS:2019mgf}, the second one to a measurement of hadronic event shapes in high-$p_{\mathrm{T}}$ multijet final states~\cite{ATLAS:2020vup}, the third one to a measurement of the Lund jet plane using charged particles~\cite{ATLAS:2020bbn} and the last one to a measurement of $b$-quark fragmentation properties in jets using the decay $B^{\pm}\rightarrow J/\psi K^{\pm}$~\cite{ATLAS:2021agf}. All these analyses use $pp$ collisions data collected with the ATLAS detector~\cite{ATLAS:2008xda} at $\sqrt{s}$ = 13~TeV at the Large Hadron Collider (LHC).

\section{Measurement of soft-drop jet observables}
\begin{figure}[t!]
\begin{center}
\includegraphics[width=0.443\textwidth]{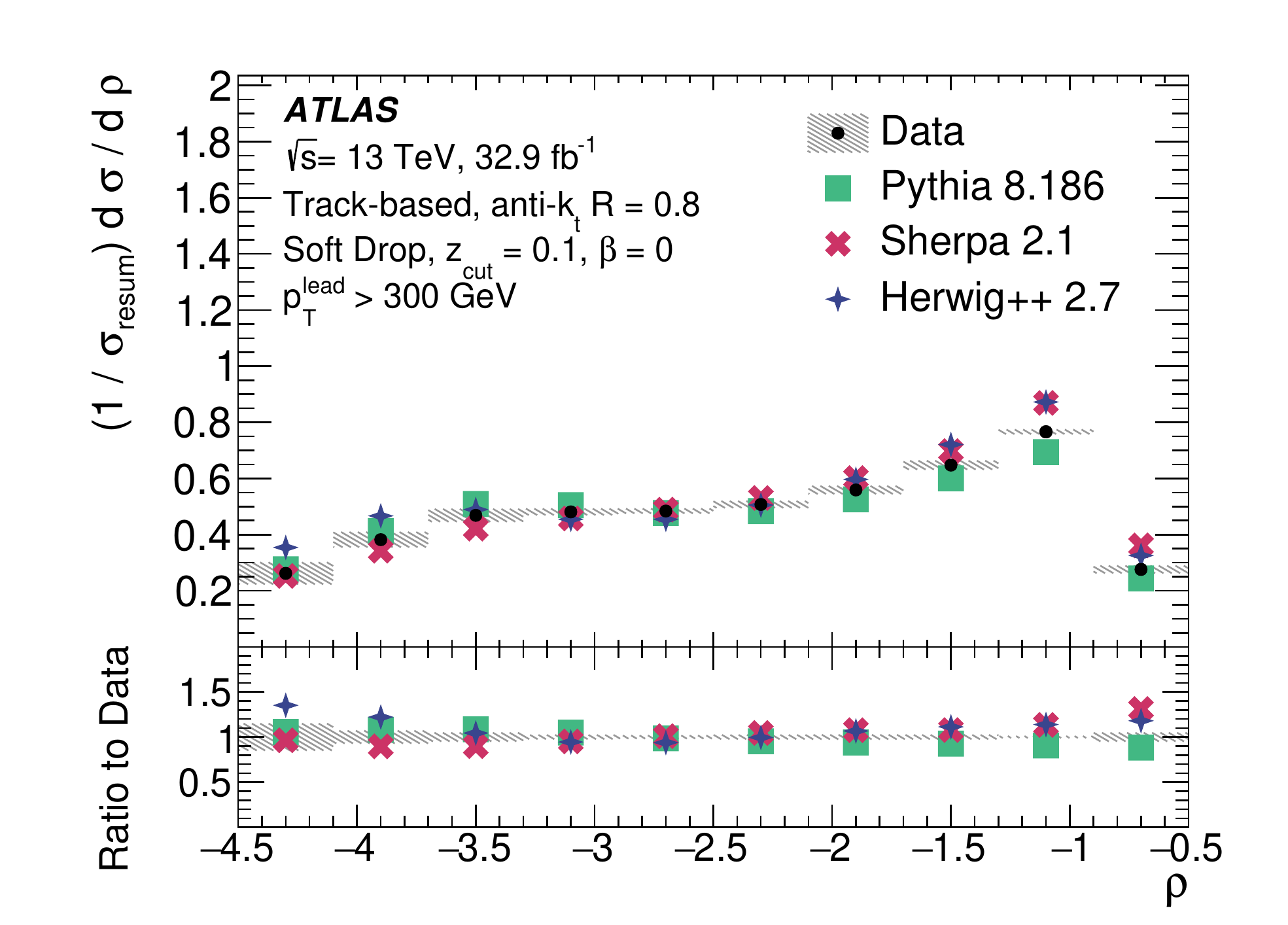}
\includegraphics[width=0.443\textwidth]{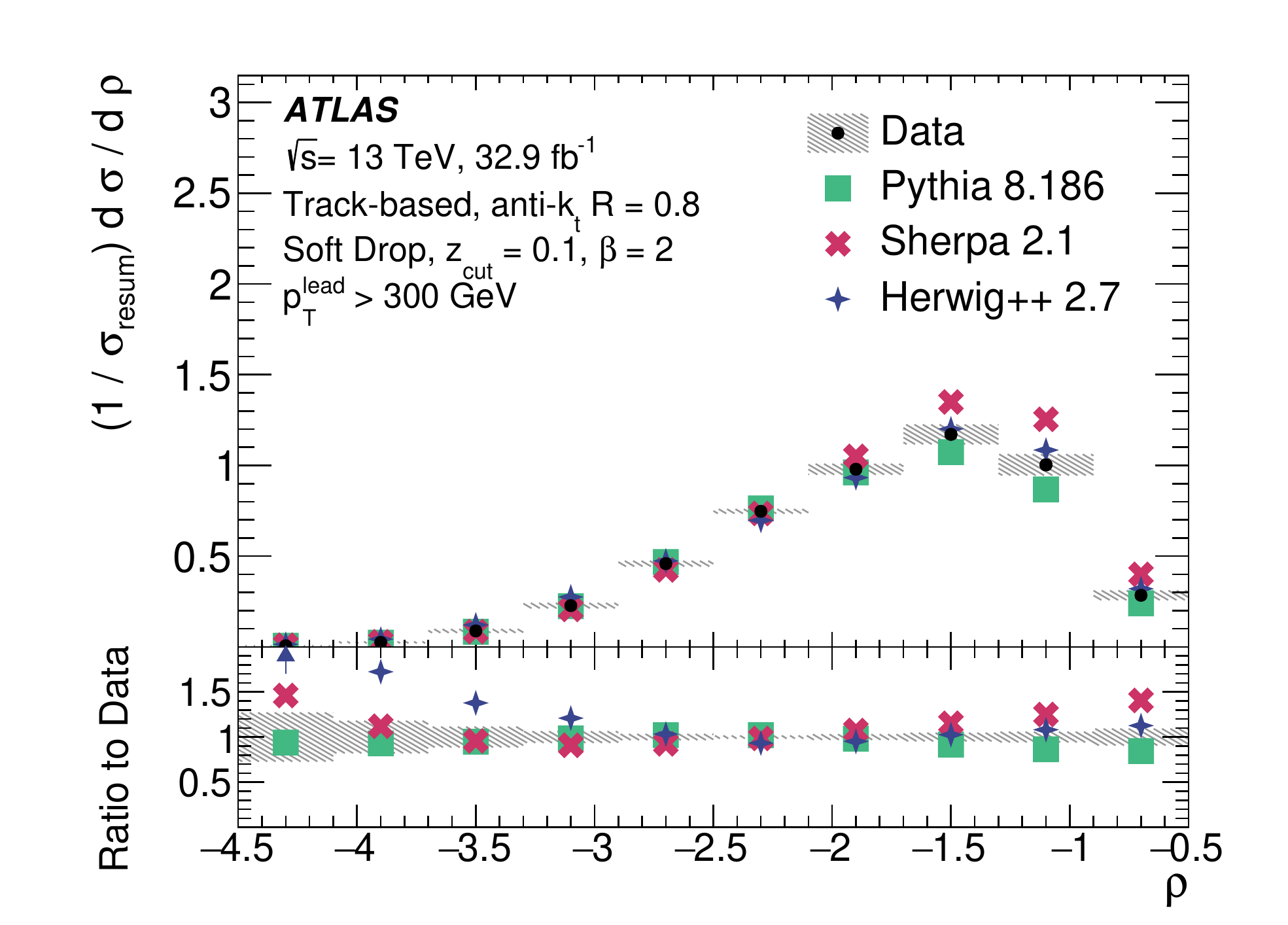}
\end{center}
\caption{Comparison of the unfolded distributions with MC predictions. The uncertainty bands include all sources of systematic uncertainties. Top left: $\rho$, $\beta=0$. Top right: $\rho$, $\beta=0$. These plots are taken from Ref.~\cite{ATLAS:2019mgf}.} 
\label{fig:softdrop}
\end{figure}

\begin{figure}[t!]
\begin{center}
\includegraphics[width=0.443\textwidth]{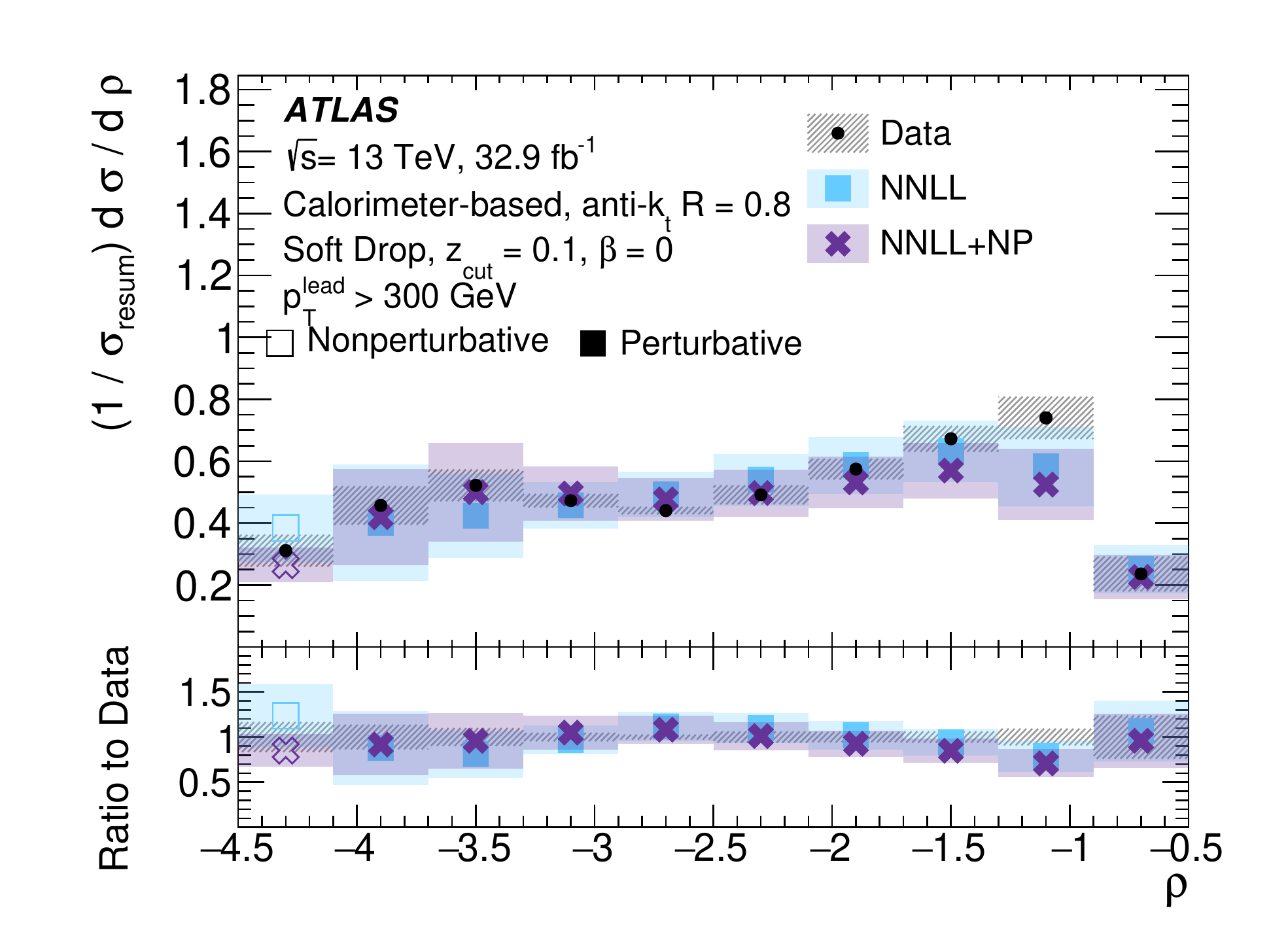}
\includegraphics[width=0.443\textwidth]{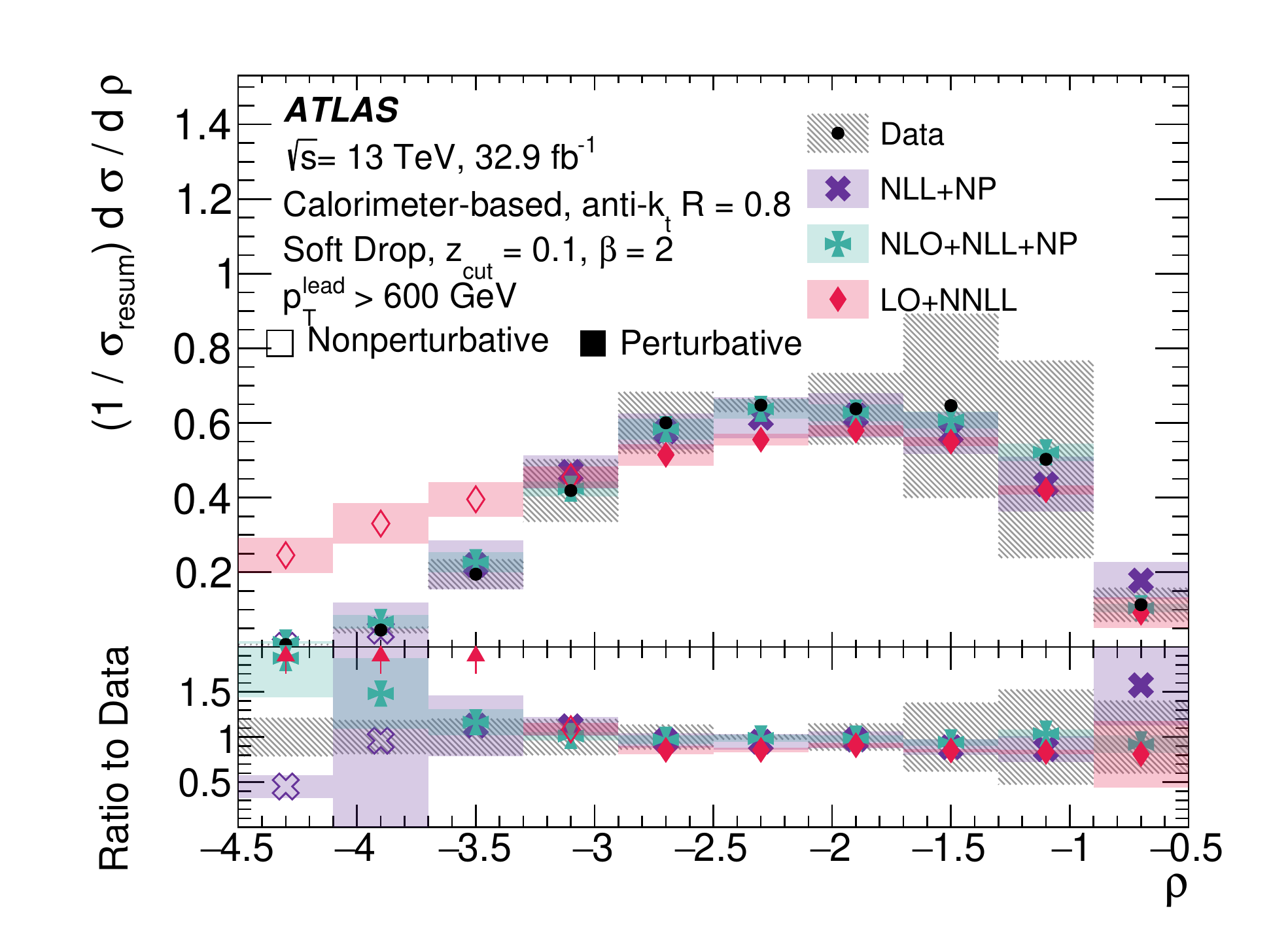}
\end{center}
\caption{Comparison of the unfolded $\rho$ distribution with different theory predictions. The open marker style indicates that non-perturbative effects on the calculation are expected to be large. ``NP'' indicates that non-perturbative corrections have been applied. The uncertainty bands include all sources of systematic uncertainties. Left: $\beta=0$. Right: $\beta=2$. These plots are taken from Ref.~\cite{ATLAS:2019mgf}.} 
\label{fig:softdrop_LL}
\end{figure}

Jet substructure quantities are measured using jets groomed with the \textit{soft-drop} grooming procedure~\cite{Larkoski:2014wba} in dijet events from data corresponding to an integrated luminosity of 32.9 fb$^{-1}$. This algorithm proceeds as follows. After a jet is clustered using any algorithm, its constituents are reclustered using the Cambrigde-Aachen (C/A) algorithm~\cite{Dokshitzer:1997in,Wobisch:1998wt}, which iteratively clusters the closest constituents in azimuth and rapidity. Then, the last step of the C/A clustering algorithm is undone, breaking the jet $j$ into two subjets, namely $j_{1}$ and $j_{2}$, which are used to evaluate the soft-drop condition:
\begin{equation}
\dfrac{\min(p_{\mathrm{T},j_{1}},p_{\mathrm{T},j_{2}})}{p_{\mathrm{T},j_{1}} + p_{\mathrm{T},j_{2}}} > \left(\dfrac{\Delta R_{12}}{R}\right)^{\beta},
\end{equation}
where $\Delta R_{12}$ is the distance between the two subjets, $R$ represents the jet radius and $p_{\mathrm{T},j_{i}}$ is the transverse momentum of the subjet $j_{i}$. The parameters $\beta$ and $z_{\mathrm{cut}}$ are algorithm parameters which determine the sensitivity of the algorithm to soft and wide-angle radiation. If the two subjets fail the soft-drop condition, the subjet characterised by the lower $p_{\mathrm{T}}$ is removed, and the other subjet is relabelled as $j$ and the procedure is iterated. When the soft-drop condition is satisfied, the algorithm is stopped, and the resulting jet is the soft-dropped jet.\\
This analysis presents two closely related substructure observables, which are calculated from jets after they have been groomed with the soft-drop algorithm: 
\begin{itemize}
\item the dimensionless version of the jet mass, $\rho=\log(m^{2}/p_{\mathrm{T}}^{2})$;
\item the opening angle between the two subjets that pass the soft-drop condition, $r_{g}$.
\end{itemize}
The unfolded data are compared to Monte Carlo (MC) events generated at leading order (LO) with \texttt{PYTHIA8.186}~\cite{Sjostrand:2006za,Sjostrand:2007gs}, \texttt{SHERPA2.1}~\cite{Gleisberg:2008ta,Sherpa:2019gpd} and \texttt{HERWIG++ 2.7}~\cite{Bahr:2008pv,Corcella:2000bw}, as reported in Figure~\ref{fig:softdrop}.
Several trends are visible in these results. For $\rho$, the MC predictions are mostly accurate within 10\% except for the lowest relative masses, which are dominated by non-perturbative physical effects. This becomes more visible for larger values of $\beta$, where more soft radiation is included within the jet, increasing the size of the non-perturbative effects. In addition, in the high-relative-mass region, where the effects of the fixed-order (FO) calculation are relevant, some differences between MC generators are seen. A similar trend may be seen for $r_{g}$, where the small-angle region (i.e. where non-perturbative effects are largest) shows more pronounced differences between MC generators.\\
Several calculations have been computed to predict the $\rho$ distributions, and unfolded data are compared with these predictions (more details on the predictions can be found in Ref.~\cite{ATLAS:2019mgf}), as shown in Figure~\ref{fig:softdrop_LL}. LO+next-to-next-to-leading-logarithm (NNLL) and next-to-leading-order+next-to-leading-logarithm (NLO+NLL) calculations are able to model the data in the resummation region ($-3\lesssim\rho\lesssim -1$), with the NLO+NLL calculation providing an accurate description of the data for high values of $\rho$. We can also observe how, in the region where the FO effects are dominant, the LO+NNLL and NNLL calculations do not model data well. This behaviour is expected, since the calculations do not include terms beyond LO at Matrix-Element (ME) level.

\section{Measurement of hadronic event shapes}
Event shapes~\cite{Banfi:2004nk,Banfi:2010xy} are a class of observables that describe the dynamics of energy flows in multijet final states, they are usually defined to be infrared and collinear safe and they are sensitive to different aspects of the theoretical description of strong-interaction processes. For example, hard, wide-angle radiation is studied by investigating the tails of these distributions, while other regions of the event-shape distributions provide information about anisotropic, back-to-back configurations, which are sensitive to the details of the resummation of soft logarithms in the theoretical predictions.\\
The dataset used in this analysis comprises the 2015-2018 data taking period, corresponding to an integrated luminosity of 139 fb$^{-1}$. In this paper, several event-shape variable are presented. For each event, the thrust axis $\hat{n}_{\mathrm{T}}$ is defined as the direction with respect to the jet momentum $p_{\mathrm{T}}$ is maximised~\cite{Brandt:1964sa,Farhi:1977sg}. The transverse thrust $T_{\perp}$ and its minor component $T_{\mathrm{m}}$ can be expressed as:
\begin{equation}
T_{\perp} = \dfrac{\sum_{i}|\vec{p}_{\mathrm{T},i}\cdot\hat{n}_{\mathrm{T}}|}{\sum_{i}|\vec{p}_{\mathrm{T},i}|}; \;\;\;\;\; T_{\mathrm{m}} = \dfrac{\sum_{i}|\vec{p}_{\mathrm{T},i}\times\hat{n}_{\mathrm{T}}|}{\sum_{i}|\vec{p}_{\mathrm{T},i}|},
\end{equation}
where the index $i$ runs over all jets in the event. These two quantities are useful to define $\tau_{\perp}=1-T_{\perp}$.\\
Several MC samples were used for this analysis, and they were produced using \texttt{PYTHIA8.235}~\cite{Sjostrand:2014zea}, \texttt{SHERPA2.1}~\cite{Gleisberg:2008ta,Sherpa:2019gpd}, \texttt{HERWIG7.1.3}~\cite{Bellm:2017bvx} and \texttt{MadGraph5{\_}aMC@NLO 2.3.3}~\cite{Alwall:2014hca}, together with \texttt{PYTHIA8.212}~\cite{Sjostrand:2014zea} (hereafter referred to as \texttt{MG5{\_}aMC}). Unfolded data are compared to the above-mentioned MC predictions in various bins of the jet multiplicity, $n^{\mathrm{jet}}$ ($=$ 2, 3, 4, 5 and $\geq$ 6), and the scalar sum of transverse momenta of the two leading jets, $H_{\mathrm{T}2}=p_{\mathrm{T1}}+p_{\mathrm{T2}}$ (1~TeV $<H_{\mathrm{T}2}<$ 1.5~TeV, 1.5~TeV $<H_{\mathrm{T}2}<$ 2.0~TeV and $H_{\mathrm{T}2}>$ 2.0~TeV).\\
\begin{figure}[t!]
\begin{center}
\includegraphics[width=0.8\textwidth]{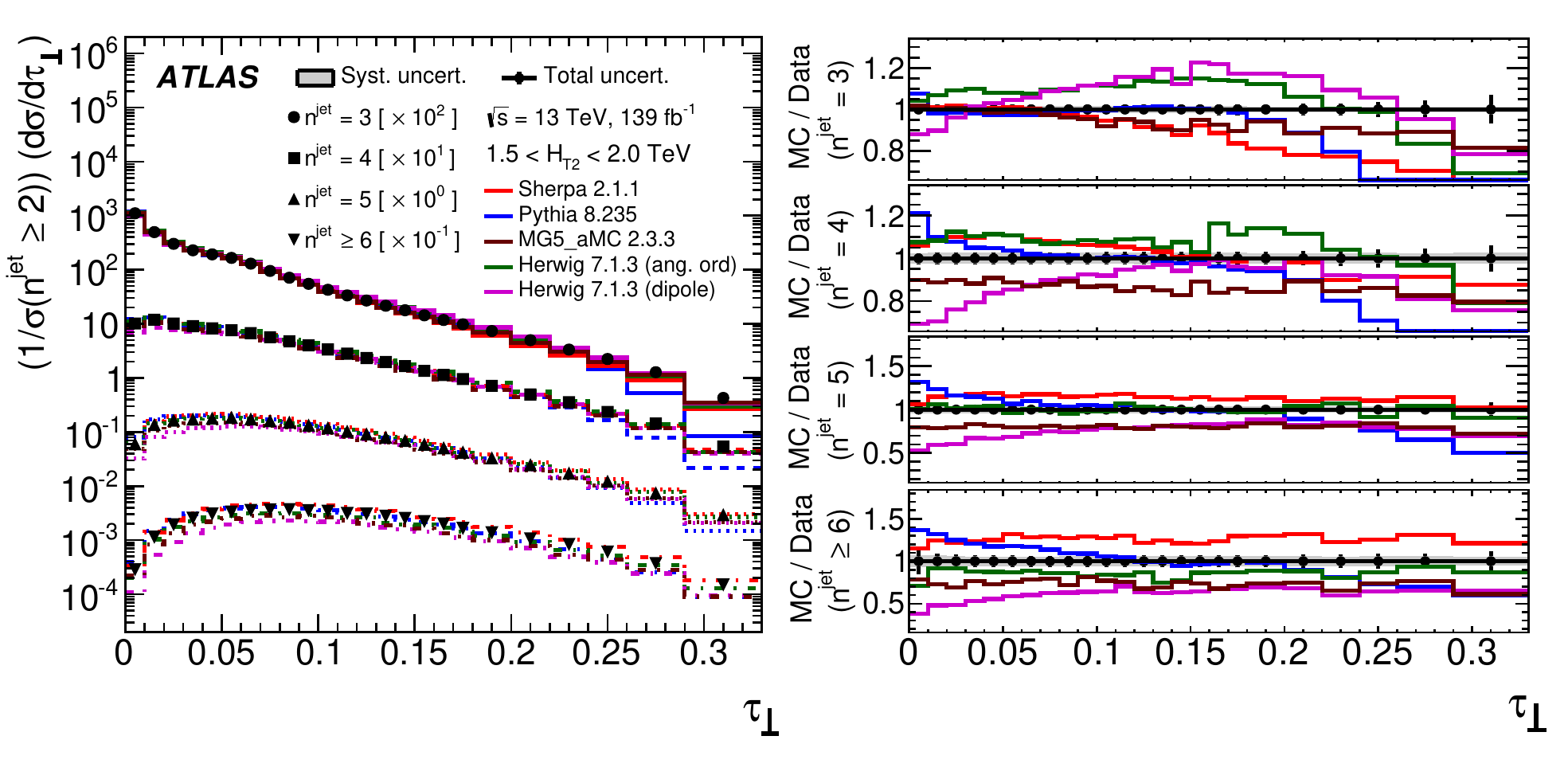}
\includegraphics[width=0.8\textwidth]{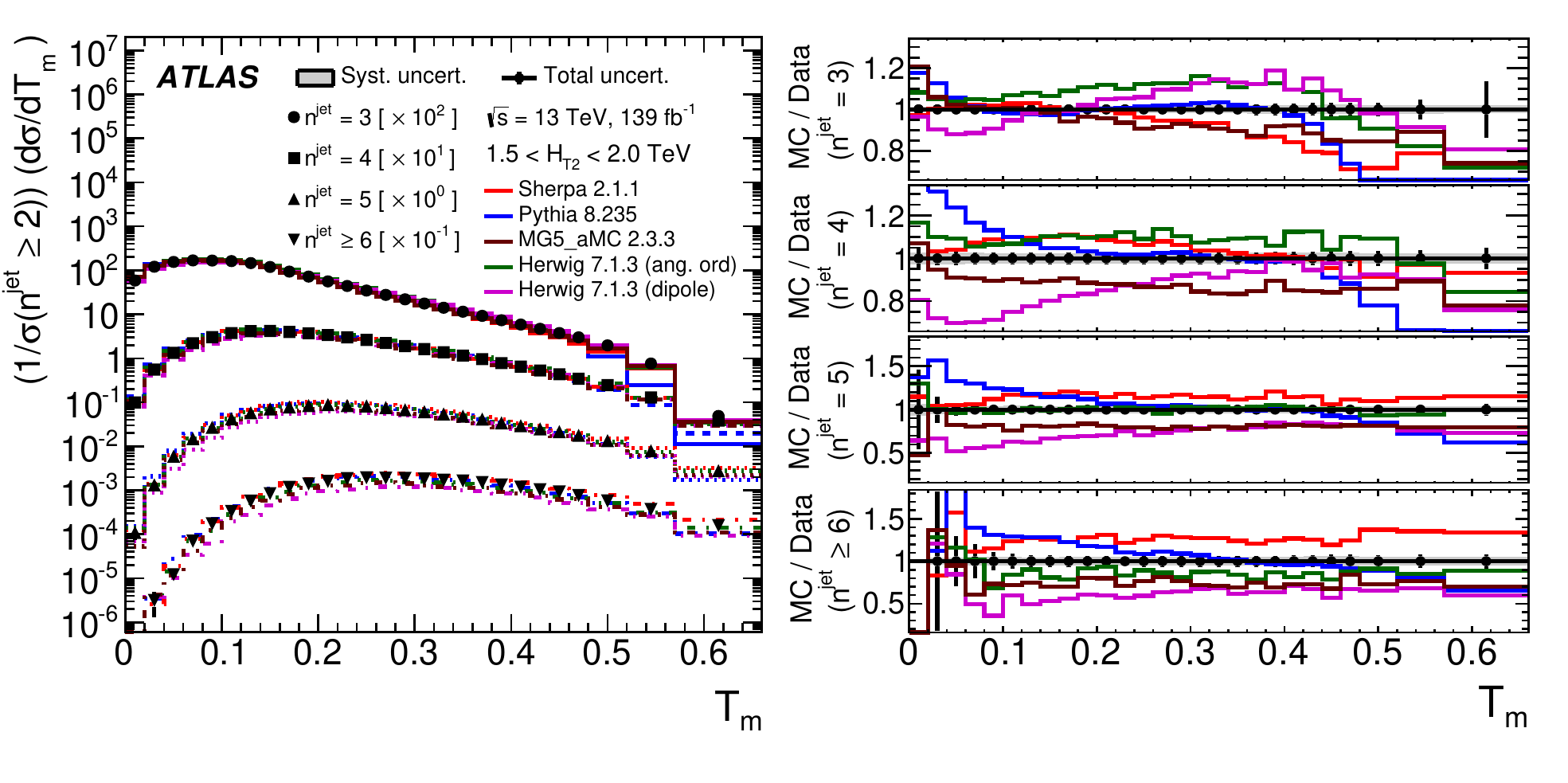}
\end{center}
\caption{Comparison between data and MC simulation for different jet multiplicities in the 1.5~TeV $<H_{\mathrm{T}2}<$ 2.0~TeV bin. The right panels show the ratios between the MC and the data distributions. The error bars show the total uncertainty (statistical and systematic added in quadrature) and the grey bands in the right panels show the systematic uncertainty. Top: normalised cross section as a function of $\tau_{\perp}$. Bottom: normalised cross section as a function of $T_{\mathrm{m}}$. These plots are taken from Ref.~\cite{ATLAS:2020vup}.} 
\label{fig:eventshape}
\end{figure}
The normalised cross section as a function of $\tau_{\perp}$ and $T_{\mathrm{m}}$ is shown in Figure~\ref{fig:eventshape}. The MC simulations tend to underestimate the data in the intermediate region of $\tau_{\perp}$ for low jet multiplicities, while the measurements are underestimated by all MC predictions at high $\tau_{\perp}$ values. The shape of the distributions tends to agree with data for larger $n^{\mathrm{jet}}$. \texttt{HERWIG7} prediction based on dipole shower highly underestimates the ATLAS data at low values of $\tau_{\perp}$, whereas the measurements are overestimated by \texttt{PYTHIA8} in such region. Very similar conclusions can be drawn looking at the normalised cross section as a function of $T_{\mathrm{m}}$. \texttt{Sherpa} simulations predict fewer isotropic events than in data, while the \texttt{MG5{\_}aMC} predictions are closer to the measurements. As regards the $H_{\mathrm{T}2}$-dependence of the depicted results, it has been found that there are more isotropic events at low energies, with increasing alignment of jets with the thrust jet axis for higher energy scales.\\
In summary, none of the MC predictions provide a good description of the ATLAS measurements in all the regions of the phase space. \texttt{HERWIG7} and \texttt{MG5{\_}aMC} computations are closer to data (discrepancies up to 1020\%), remarking the limited ability of parton shower (PS) models to simulate hard and wide angle radiation and further emphasising that the addition of $2\rightarrow3$ processes in the ME allows to improve the description of measured data.

\section{Measurement of the Lund jet plane}
\begin{figure}[t!]
\begin{center}
\includegraphics[width=0.43\textwidth]{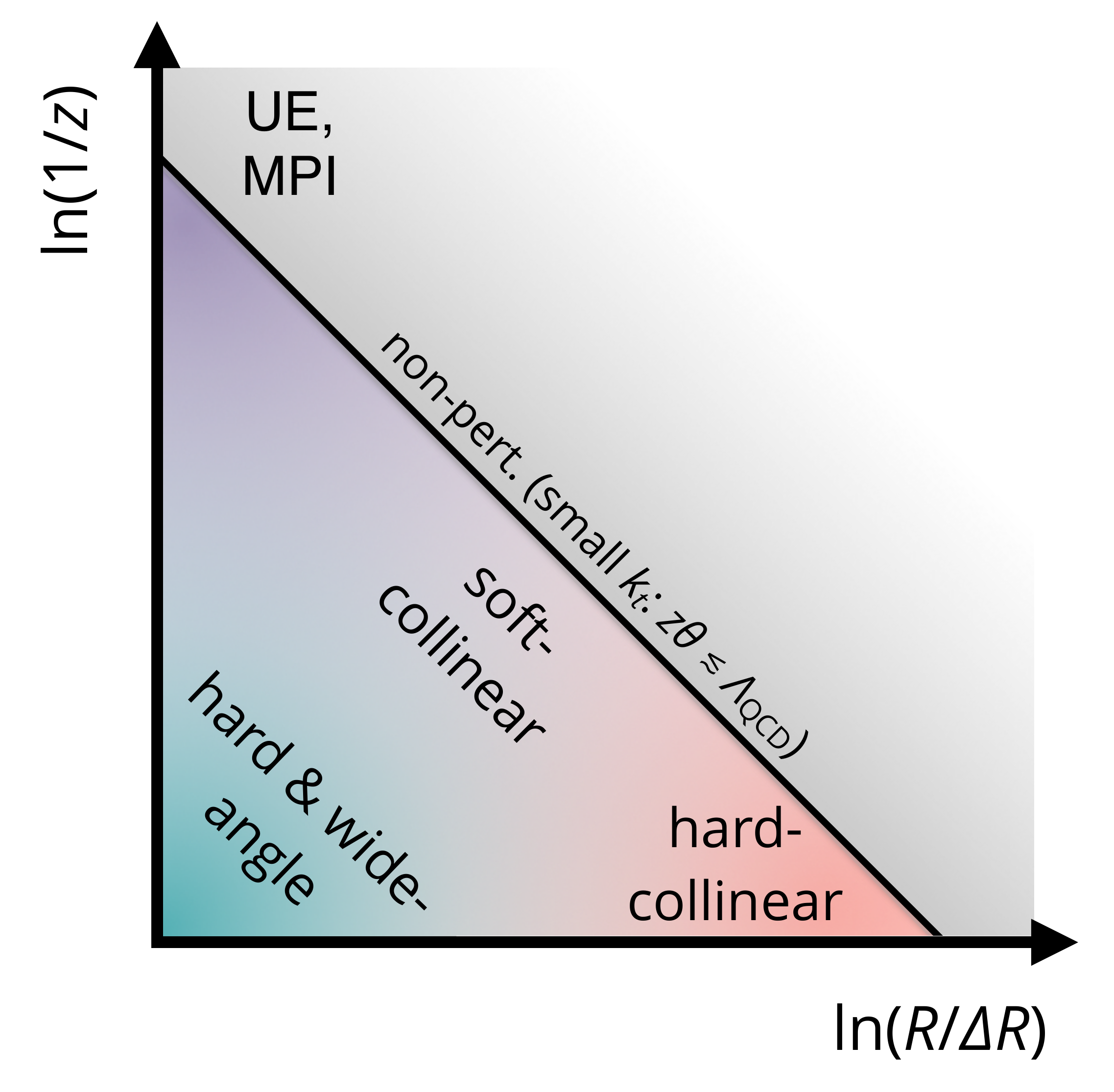}
\includegraphics[width=0.43\textwidth]{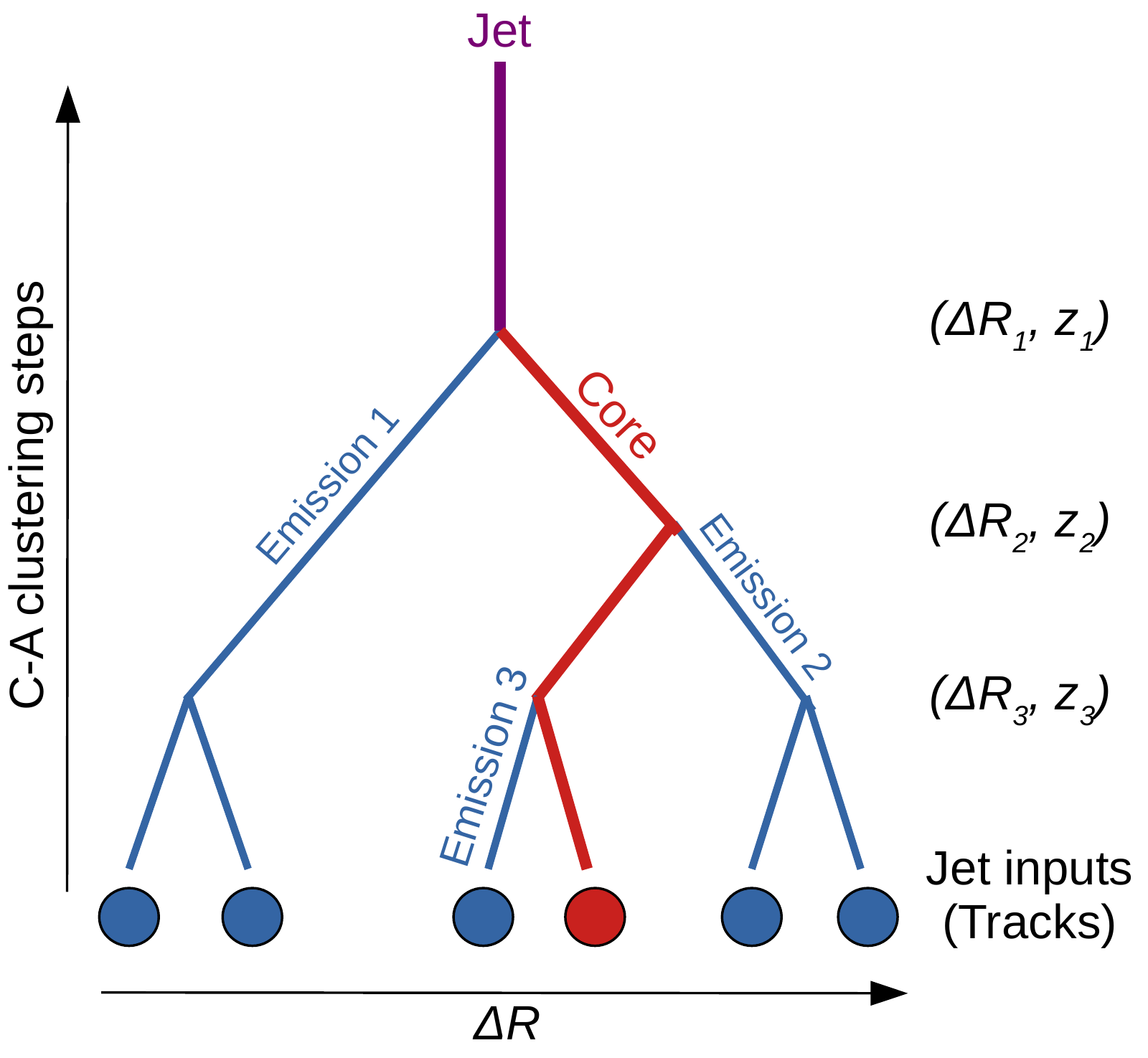}
\end{center}
\caption{Schematic representation of the LJP. The left-hand side plots is taken from Ref.~\cite{ATLAS:2020bbn}.} 
\label{fig:LJP}
\end{figure}
\begin{figure}[t!]
\begin{center}
\includegraphics[width=0.443\textwidth]{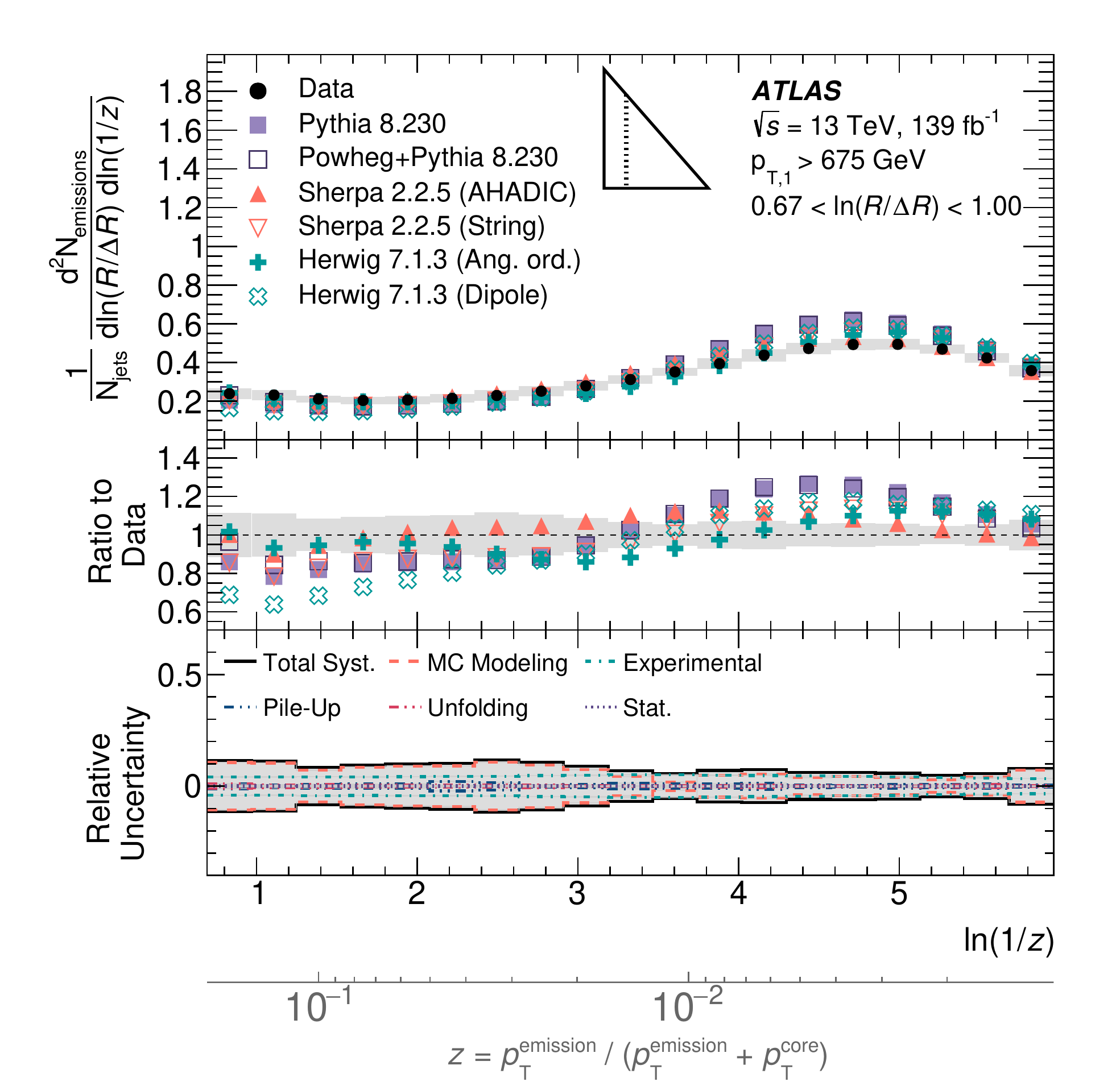}
\includegraphics[width=0.443\textwidth]{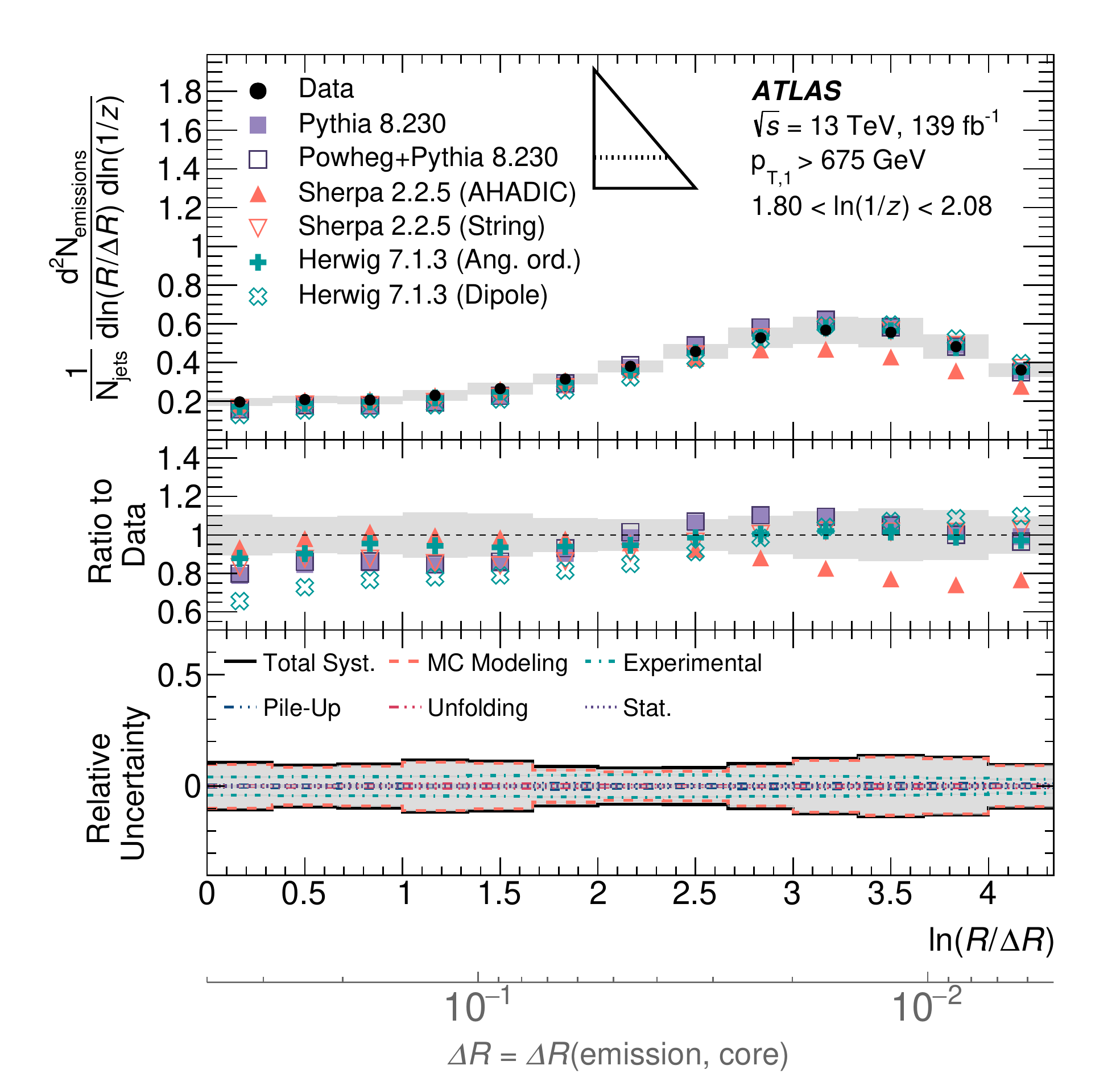}
\end{center}
\caption{Representative horizontal and vertical slices through the LJP. Unfolded data are compared with particle-level simulation from several MC generators. The uncertainty band includes all sources of systematic and statistical uncertainty. The inset triangle illustrates which slice of the plane is depicted Left: 0.67 $<\ln(R/\Delta R)<$ 1.0. Right: 1.80 $<\ln(1/z)<$ 2.08. These plots are taken from Ref.~\cite{ATLAS:2020bbn}.} 
\label{fig:LJP_data}
\end{figure}

The Lund plane is a powerful representation for providing insight into jet substructure. A recent proposal~\cite{Dreyer:2018nbf} describes a method to construct an observable analog of the Lund plane using jets, which captures the salient features of this representation. Jets are formed using clustering algorithms that sequentially combine pairs of proto-jets starting from the initial set of constituents~\cite{Salam:2010nqg}. In this proposal, a jet's constituents are reclustered using the C/A algorithm~\cite{Dokshitzer:1997in,Wobisch:1998wt}. Then, the C/A history is reversed, and each jet is declustered, starting from the hardest proto-jet. The Lund plane can be approximated by using the harder (softer) proto-jet to present the core (emission) in the original theoretical depiction. For each proto-jet pair, at each step in the C/A declustering sequence, an entry is made in the primary Lund Jet plane (LJP) through the observables $\ln(1/z)$ and $\ln(R/\Delta R)$, with
\begin{equation}
z=\dfrac{p_{\mathrm{T}}^{\mathrm{emission}}}{p_{\mathrm{T}}^{\mathrm{emission}}+p_{\mathrm{T}}^{\mathrm{core}}} \;\; \mathrm{and} \;\; \Delta R^{2} = (y_{\mathrm{emission}} - y_{\mathrm{core}})^{2} + (\phi_{\mathrm{emission}} - \phi_{\mathrm{core}})^{2}.
\end{equation}
A schematic representation of the LJP can be found in Figure~\ref{fig:LJP}.\\
This measurement is conducted using the full Run 2 statistics, for an integrated luminosity of 139 fb$^{-1}$. To perform the data unfolding, several samples of dijet events were simulated. \texttt{PYTHIA8.186}  \cite{Sjostrand:2006za,Sjostrand:2007gs} was used for simulating the nominal sample. Additional samples were simulated using NLO MEs from \texttt{POWHEG}~\cite{Nason:2004rx,Frixione:2007vw,Alioli:2010xa,Alioli:2010xd} and \texttt{Sherpa2.2.5}~\cite{Sherpa:2019gpd} or \texttt{HERWIG 7.1.3}~\cite{Bellm:2017bvx}.\\
The data from two seleceted slices of the LJP, together with the breakdown of the major systematic uncertainties, are shown in Figure~\ref{fig:LJP_data}. ATLAS data and several MC predictions are compared. It is visible how the \texttt{Herwig7.1.3} angle-ordered prediction provides the best description across most of the plane, while any prediction describes the data accurately in all the regions. The differences in the hadronization algorithms implemented in \texttt{Sherpa2.2.5} are particularly visible at the transition between perturbtive and non-perturbative regions of the plane. The \texttt{POWHEG+PYTHIA} and \texttt{PYTHIA} predictions only differ significantly for hard and wide-angle perturbative emissions, where ME corrections are relevant.

\section{Measurement of $b$-quark fragmentation properties}
\begin{figure}[H]
\begin{center}
\includegraphics[width=0.443\textwidth]{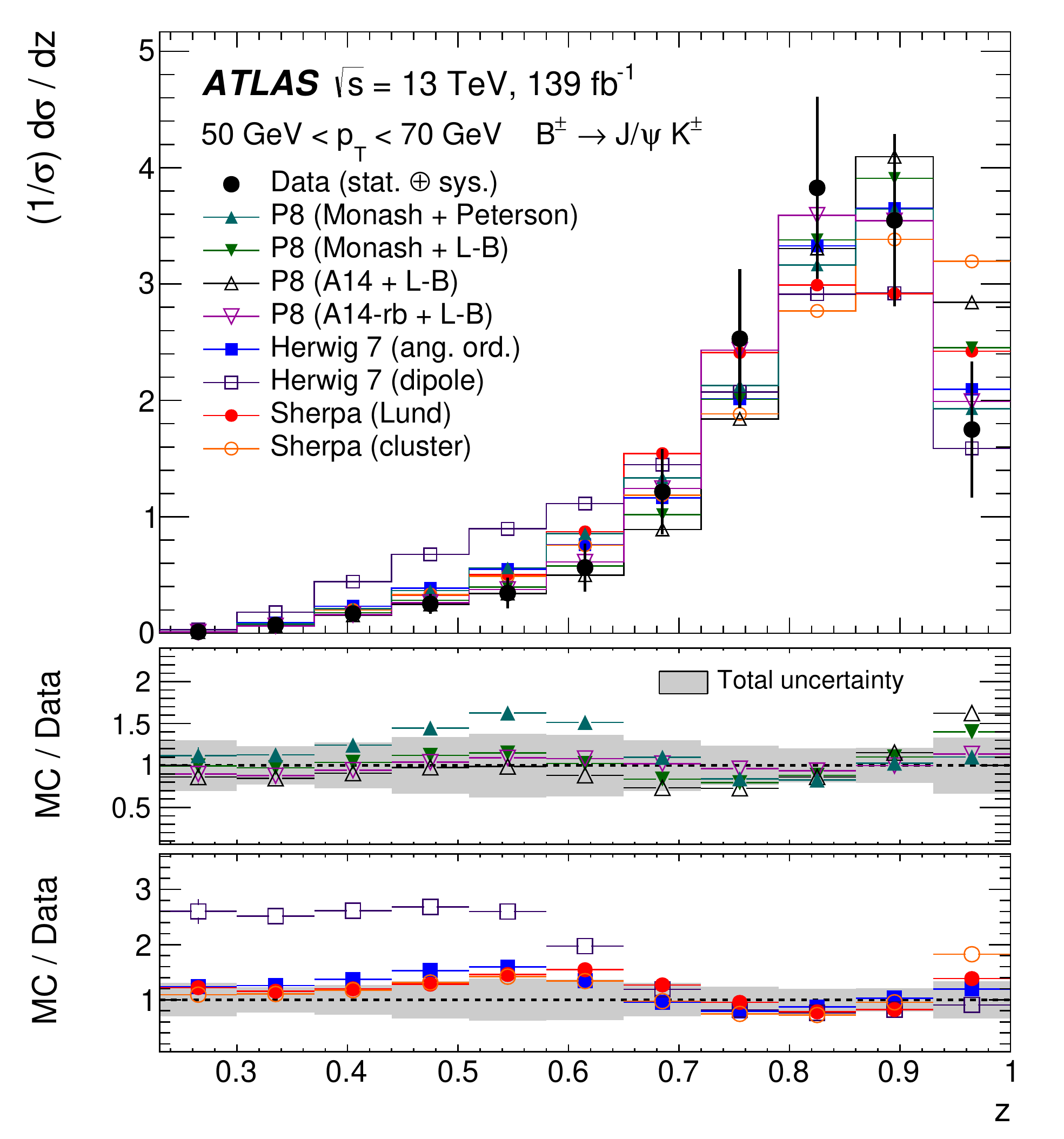}
\includegraphics[width=0.443\textwidth]{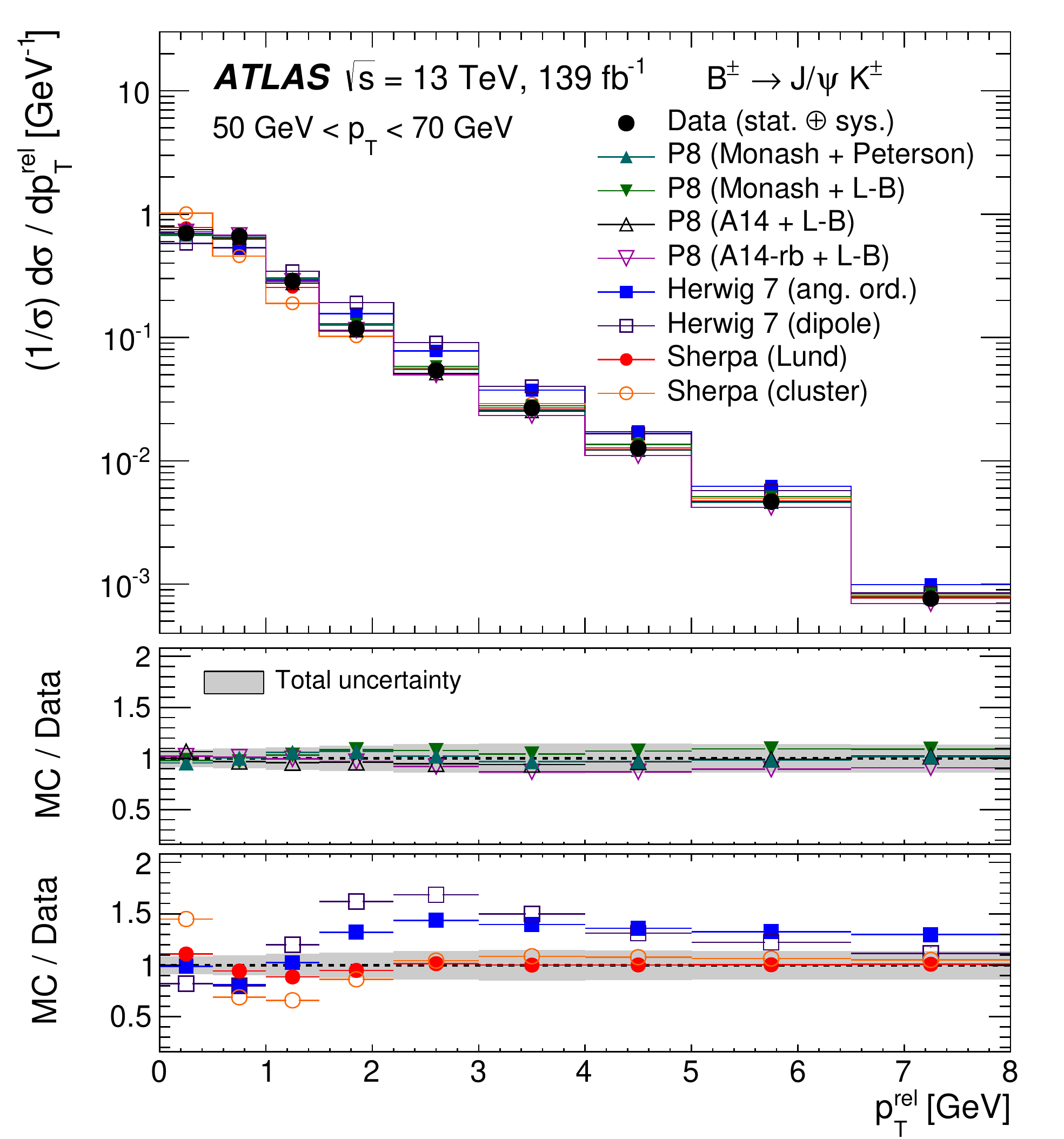}
\includegraphics[width=0.443\textwidth]{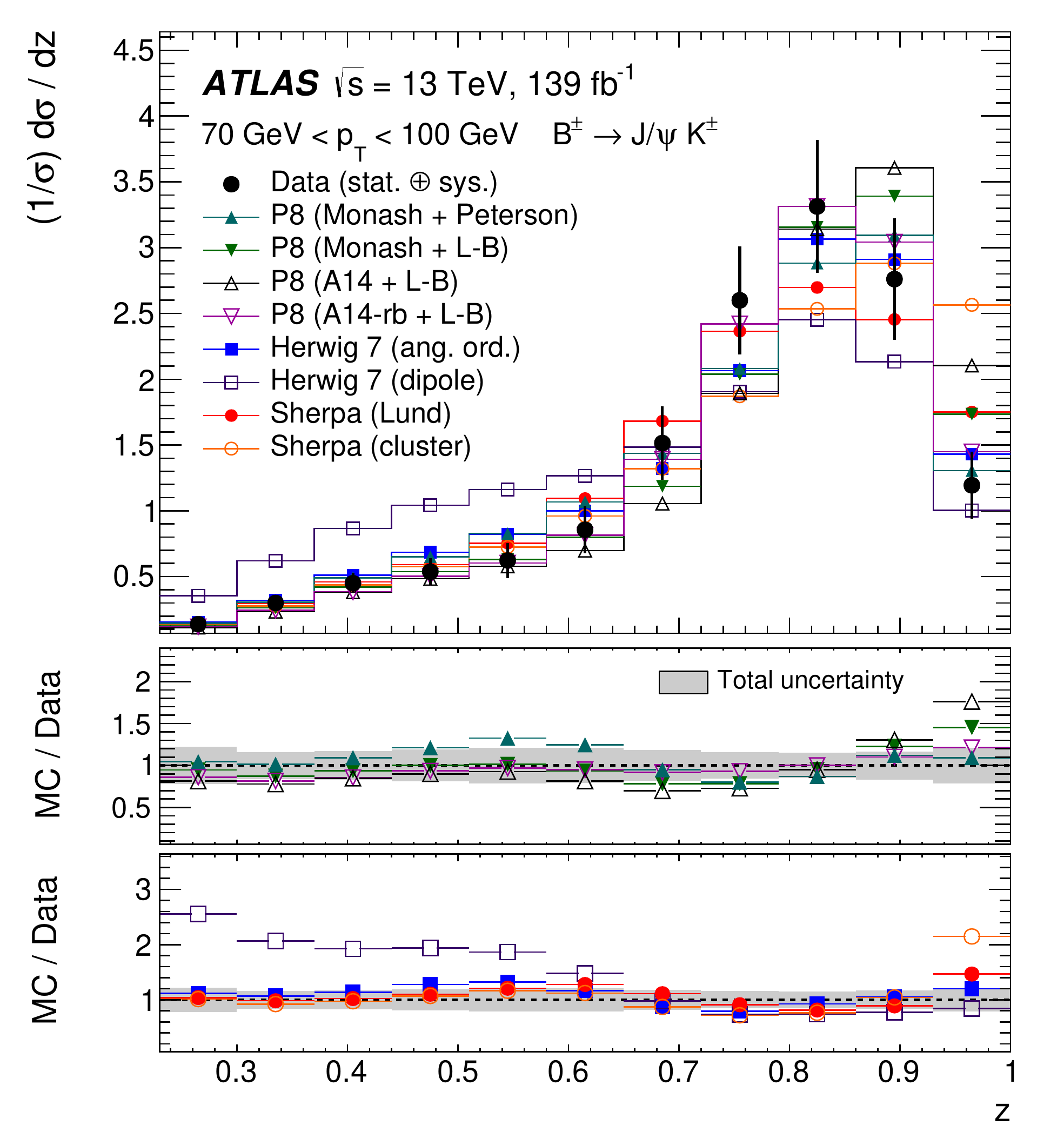}
\includegraphics[width=0.443\textwidth]{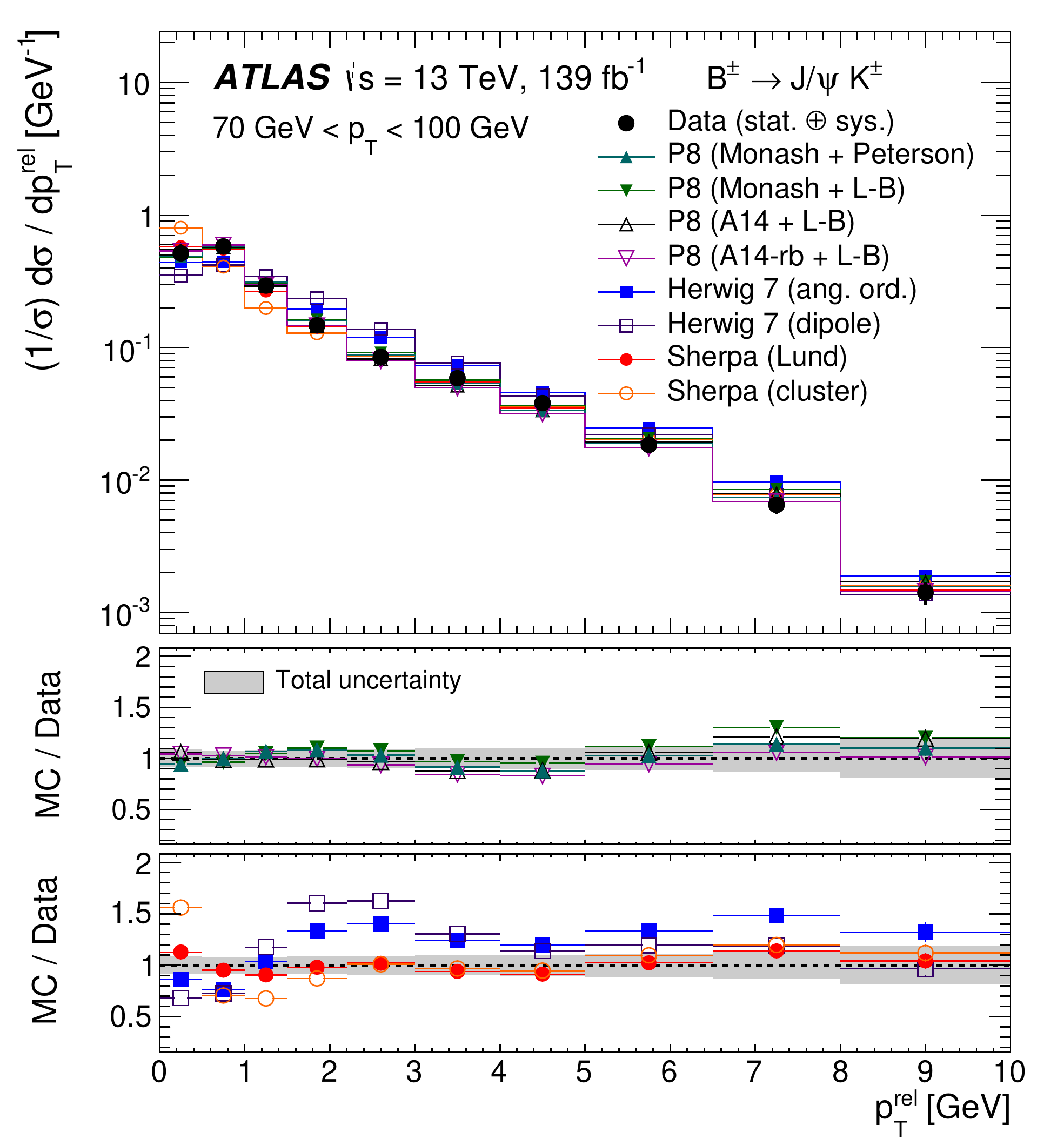}
\includegraphics[width=0.443\textwidth]{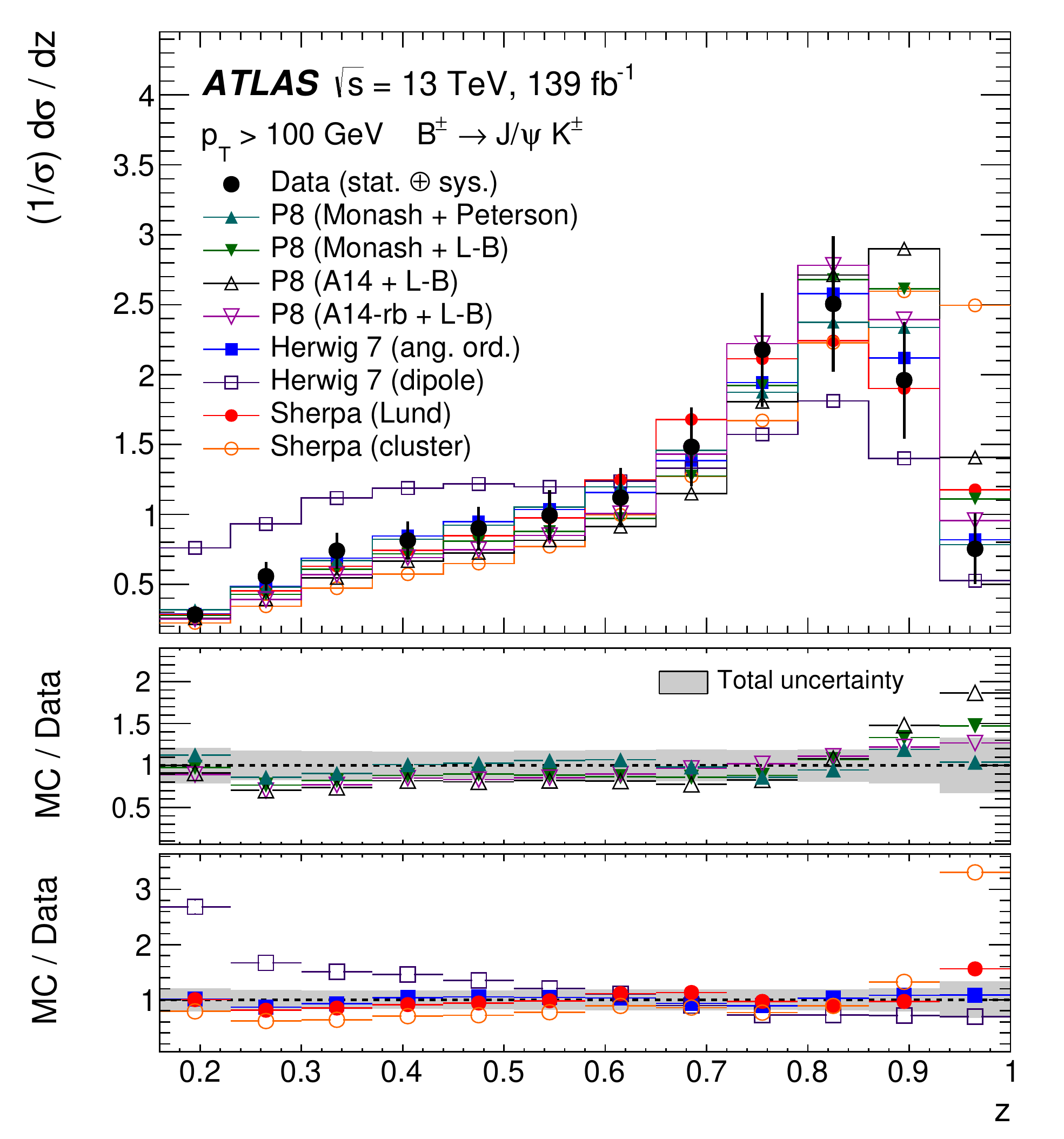}
\includegraphics[width=0.443\textwidth]{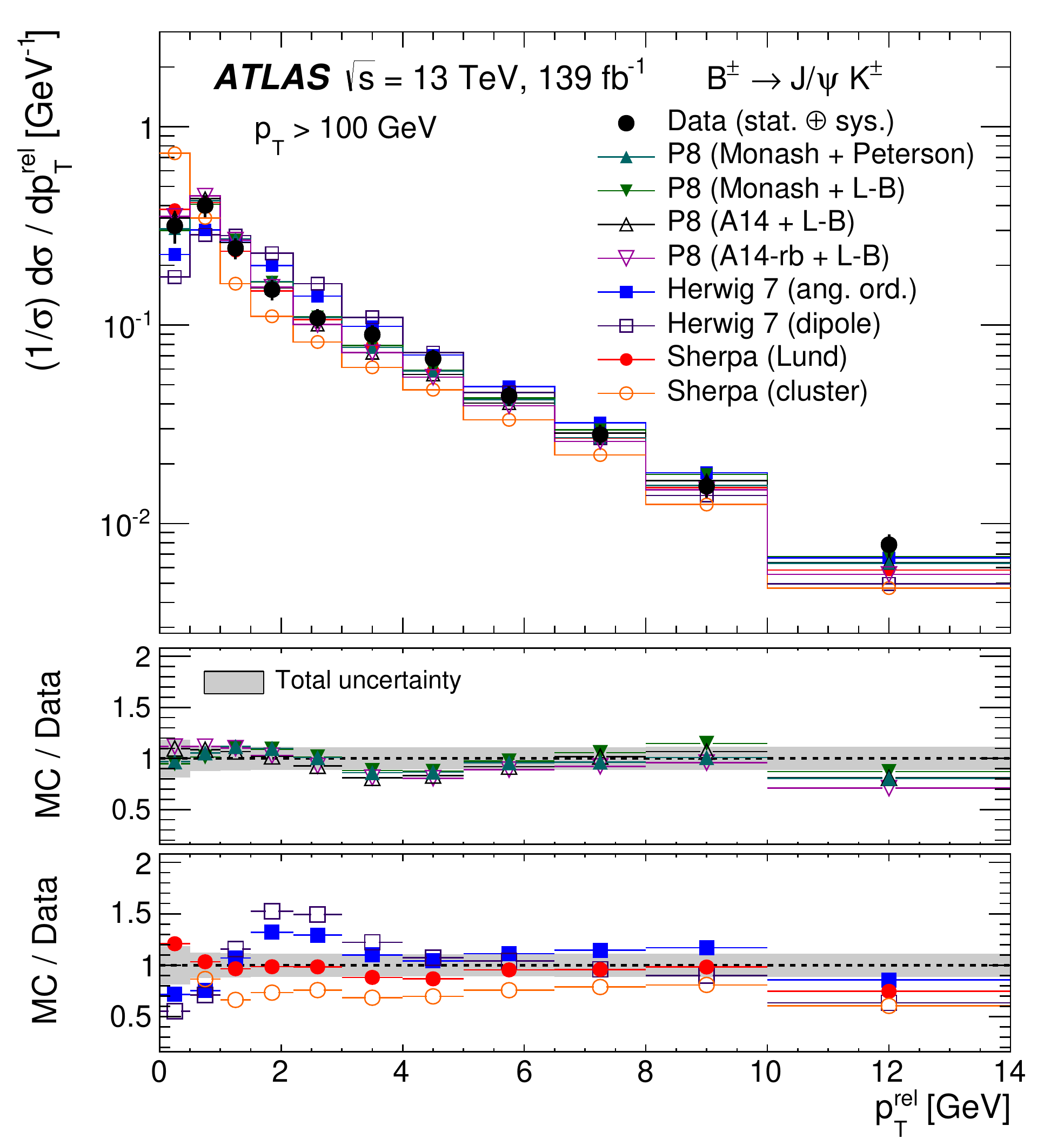}
\end{center}
\caption{Distributions of the longitudinal profile $z$ and the transverse profile $p_{\mathrm{T}}^{\mathrm{rel}}$ together with different predictions from \texttt{PYTHIA8}, \texttt{SHERPA} and \texttt{HERWIG 7}. The vertical error bars represent the total experimental uncertainties. Top: 50 $<p_{\mathrm{T}}^{\mathrm{jet}}<$ 70~GeV. Middle: 70 $<p_{\mathrm{T}}^{\mathrm{jet}}<$ 100~GeV. Bottom: $p_{\mathrm{T}}^{\mathrm{jet}}>$ 100~GeV. These plots are taken from Ref.~\cite{ATLAS:2021agf}.} 
\label{fig:bfrag}
\end{figure}

The fragmentation of heavy quarks is a crucial aspect of Quantum ChromoDynamics (QCD). Detailed studies and precision measurements of the heavy-quark fragmentation properties allow a deeper understanding of QCD. The MC predictions used at the LHC are tuned to describe the measurements in $e^{+}e^{-}$ collisions at relatively low $\sqrt{s}$. Therefore, new measurements of $b$-quark fragmentation can be used to improve MC simulations at LHC energy scales.\\
This analysis presents a measurement of $b$-quark fragmentation into $B^{\pm}$ mesons and it uses the full Run 2 data set, corresponding to an integrated luminosity of 139 fb$^{-1}$. The $B^{\pm}$ mesons are then recostructed via the $B^{\pm}\rightarrow J/\psi K^{\pm}\rightarrow\mu^{+}\mu^{-}K^{\pm}$ decay chain. After the matching between the jet and the reconstructed $B$ meson, two variables of interest are built as follows:
\begin{equation}
z=\dfrac{\vec{p}_{B}\cdot\vec{p}_{j}}{|\vec{p}_{j}|^{2}} \;\; \mathrm{and} \;\; p_{\mathrm{T}}^{\mathrm{rel}}=\dfrac{\vec{p}_{B}\times\vec{p}_{j}}{|\vec{p}_{j}|},
\end{equation}
where $\vec{p}_{B}$ is the three-momentum of the $B$ hadron and $\vec{p}_{j}$ is the jet three-momentum. The measurement is performed in three different intervals of the jet transverse momentum, namely: 50 $<p_{\mathrm{T}}^{\mathrm{jet}}<$ 70~GeV, 70 $<p_{\mathrm{T}}^{\mathrm{jet}}<$ 100~GeV and $p_{\mathrm{T}}^{\mathrm{jet}}>$ 100~GeV.\\
Several different models of multijet production are used. These samples have been generated using \texttt{SHERPA2.2.5}~\cite{Sherpa:2019gpd}, \texttt{PYTHIA8.240}~\cite{Sjostrand:2006za,Sjostrand:2007gs} and \texttt{HERWIG7.2.1}~\cite{Bellm:2017bvx}, with substantial differences in the ME calculations, as well as PS algorithms and hadronisation models. The decays of the $B$ mesons were modelled using \texttt{EVTGEN1.6.0} code~\cite{Lange:2001uf} for all the above-mentioned samples.\\
These predictions are then comprared with the particle-level results, as shown in Figure~\ref{fig:bfrag}, where the longitudinal ($z$) and transverse  ($p_{\mathrm{T}}^{\mathrm{rel}}$) profiles for each $p_{\mathrm{T}}$ bin are reported. The results show important differences between the low and high $p_{\mathrm{T}}$ bins. Particularly, the lower tails of the $z$ distributions contain a larger fraction of the high-$p_{\mathrm{T}}$ data due to the larger probability of having gluon splitting, $g\rightarrow b\bar{b}$ when considering high values of the jet $p_{\mathrm{T}}$.\\
The \texttt{SHERPA} predictions give a reasonable description of the $z$ distributions in the low and medium $p_{\mathrm{T}}$ bins, although they differ from data for very high values of $z$. They also show large discrepancies for low values of $p_{\mathrm{T}}^{\mathrm{rel}}$, which increase when moving towards higher bins of the jet $p_{\mathrm{T}}$. All the various \texttt{PYTHIA8} samples provide a good description of the $z$ and $p_{\mathrm{T}}^{\mathrm{rel}}$ distributions, being compatible with data within the systematic uncertainties across the different jet-$p_{\mathrm{T}}$ bins. The results for the longitudinal profile show reasonable agreement with the \texttt{HERWIG7} prediction with the angle-ordered PS, while large discrepancies are observed with the dipole parton shower. Due to the larger gluon splitting fractions, the \texttt{HERWIG7} sample with the dipole PS significantly overestimates the data in the tails of the $p_{\mathrm{T}}^{\mathrm{rel}}$ distributions at low $p_{\mathrm{T}}$, while the differences are smaller with increasing $p_{\mathrm{T}}$. The Herwig angle-ordered PS gives a better description of the $p_{\mathrm{T}}^{\mathrm{rel}}$ distributions, although non-negligible discrepancies are also observed.

\section{Conclusion}
Measurements of variables probing the properties of the multijet energy flow and of the Lund Plane using charged particles, as well as a measurement of the fragmentation properties of $b$-quark initiated jets, have been presented in this proceeding. Ref.~\cite{ATLAS:2019mgf} demonstrates differences between the soft-drop jet substructure observables in their sensitivity to the quark and gluon composition of the sample, which are most pronounced for the least amount of grooming. In Ref~\cite{ATLAS:2020vup} the discrepancies between event shapes data and all the investigated MC show that further refinement of the current MC predictions is needed to describe the data in some regions, particularly at high jet multiplicities. Ref.~\cite{ATLAS:2020bbn} illustrates the ability of the Lund jet plane to isolate various physical effects, and will provide useful input to both perturbative and nonperturbative model development and tuning. Finally, Ref.~\cite{ATLAS:2021agf} provides key measurements with which to better understand the fragmentation functions of heavy quarks. As has been shown, significant differences among different MC models are observed, and also between the models and the data. Some of the discrepancies are understood to arise from poor modelling of the $g\rightarrow b\bar{b}$ splittings, to which the present analysis has substantial sensitivity. Including the present measurements in a future tune of the MC predictions may help to improve the description and reduce the theoretical uncertainties of processes where heavy-flavour quarks are present in the final state, such as top quark pair production or Higgs boson decays into heavy quark pairs.

\end{document}